# A NOVEL FRAMEWORK TO ASSESS CYBERSECURITY CAPABILITY MATURITY

*Short Paper*

Lasini Liyanage, University of Auckland, New Zealand, lliy930@aucklanduni.ac.nz

Nalin Arachchilage, RMIT University, Australia, nalin.arachchilage@rmit.edu.au

Giovanni Russello, University of Auckland, New Zealand, g.russello@auckland.ac.nz

## Abstract

*In today's rapidly evolving digital landscape, organisations face escalating cyber threats that can disrupt operations, compromise sensitive data, and inflict financial and reputational harm. A key reason for this lies in the organisations' lack of a clear understanding of their cybersecurity capabilities, leading to ineffective defences. To address this gap, Cybersecurity Capability Maturity Models (CCMMs) provide a systematic approach to assessing and enhancing an organisation's cybersecurity posture by focusing on capability maturity rather than merely implementing controls. However, their limitations, such as rigid structures, one-size-fits-all approach, complexity, gaps in security scope (i.e., technological, organisational, and human aspects) and lack of quantitative metrics, hinder their effectiveness. It makes implementing CCMMs in varying contexts challenging and results in fragmented, incomprehensive assessments. Therefore, we propose a novel Cybersecurity Capability Maturity Framework that is holistic, flexible, and measurable to provide organisations with a more relevant and impactful assessment to enhance their cybersecurity posture.*

*Keywords: cybersecurity capability maturity, organisational security, cybersecurity framework, maturity model.*

## 1 Introduction

Organisations today operate in a rapidly evolving digital environment where technological advancements present both opportunities and escalating cyber threats. According to the Cyber Readiness Report 2024 (Hiscox, 2024), the average number of cyberattacks experienced per organisation rose from 63 in 2022/23 to 66 in 2023/24, highlighting the growth of cyber threats. By 2025, cyber attacks are expected to cost $10.5 trillion annually (Hiscox, 2024). Despite this growing risk, one question remains unanswered for many organisations: How capable are they of defending against these cyber attacks?

For many organisations, this uncertainty is the root of their vulnerability to cyberattacks (Liyanage et al., 2024). Without a clear understanding of their cybersecurity capabilities, they remain exposed. When organisations fail to grasp the true nature of their strengths and weaknesses, they may have unguarded systems or overlooked processes that can become the cracks through which attackers gain entry. This is where a systematic approach to analysing cybersecurity capabilities becomes essential (Aliyu et al., 2020). Organisations need a structured method to evaluate where they stand, assess the effectiveness of their existing defences, identify areas where improvement is critical, and ensure that they are investing in the right solutions (Rabii et al., 2020). Without this, they are left navigating in the dark, making decisions without a true understanding of the impact.

Cybersecurity Capability Maturity Models (CCMMs) are essential in this context, providing comprehensive frameworks for assessing and improving an organisation's cybersecurity posture (Rea-Guaman et al., 2017). Unlike general security frameworks such as the NIST Cybersecurity Framework (NIST CSF) or international standards like ISO/IEC 27001, which primarily focus on establishing baseline controls such as access management, incident response and vulnerability assessments to ensure regulatory compliance, CCMMs provide a progressive roadmap for assessing current capabilities, identifying gaps, and implementing targeted improvements (Rea-Guaman et al., 2017).

Despite the advantages provided by the CCMMs, existing CCMMs adopt a one-size-fits-all approach, failing to accommodate the diverse sizes, risk profiles, and operational needs of different organisations (Aliyu, et al., 2020; Dube and Mohanty, 2020). This often results in ineffective security measures that do not adequately protect their unique assets. Additionally, current CCMMs have gaps in their consideration of cybersecurity capabilities across technological, organisational, operational, and human aspects (i.e. holistic approach), such as employee training, collaboration and information sharing, data security, physical security, performance evaluation and business continuity planning (Liyanage et al., 2024). This narrow focus can lead to security gaps, making organisations vulnerable. For instance, neglecting employee training (human aspect) increases susceptibility to phishing attacks, while failing to plan for business continuity (operational aspect) leaves organisations exposed to prolonged disruptions after incidents. Similarly, inadequate data security (technological aspect) risks breaches, and poor collaboration (organisational aspect) hinders effective incident response (Aliyu et al., 2020).

Furthermore, the complexity and rigidity of existing CCMMs pose challenges for implementation and customisation, leading to longer adoption times and higher costs (Liyanage et al., 2024). Organisations may be discouraged from fully utilising these models, leaving gaps in their security posture. The absence of quantitative metrics in many CCMMs also hampers the objective measurement of cybersecurity effectiveness and progress (Le and Hoang, 2017), making it difficult for organisations to prioritise improvements and demonstrate compliance or secure funding for cybersecurity initiatives. These limitations collectively increase organisations' susceptibility to cyber threats and impede their ability to continuously improve their security posture.

Motivated by these challenges and building on work in Liyanage et al. (2024), this study proposes a novel Cybersecurity Capability Maturity Framework (CCMF), as shown in Figure 1, based on a Design Science Research (DSR) approach. The framework is holistic, flexible, and measurable, providing organisations with a more relevant and impactful assessment to enhance their cybersecurity posture. To achieve this, we seek to answer the following research question:

**RQ1: What are the core components of a tailored and comprehensive CCMF that effectively assesses cybersecurity capability maturity in organisations?**

To address this research question, our proposed CCMF identifies and incorporates key components necessary for a comprehensive maturity assessment. These components include *(1) customizable assessment criteria* that allow organisations to tailor the framework to their specific security priorities (e.g., protecting sensitive customer data or securing critical infrastructure), operational context, and available resources (e.g., budget, staff expertise, and technological assets), *(2) holistic capabilities* that integrate technological, organisational, operational, and human factors to ensure a well-rounded security posture, and *(3) quantitative metrics* to objectively measure the effectiveness of capabilities and track improvements over time. These components are incorporated to address the shortcomings of existing CCMMs (i.e. one-size-fits-all approach, lack of holistic approach and quantitative metrics), providing a structured and flexible approach to understanding and improving an organisation's cybersecurity maturity. To facilitate adoption and ease of use, our framework is implemented as an interactive, free, web-based tool, allowing organisations to intuitively assess their maturity levels and visualise their cybersecurity posture across various domains.

## 2      Background and Related Work

Early efforts to assess overall cybersecurity capability maturity in organisations relied on Generic CCMMs such as C2M2, Adaptive CCMM, and CM[2] (Rabii et al., 2020). These models provide a standardised approach for evaluating an organisation's security posture (Alayo et al., 2021). However, Ali et al. (2022) argue that these models were complex and often led to a convoluted assessment process, making it difficult for organisations with limited resources to navigate the detailed requirements and criteria. This complexity also increased the likelihood of misinterpretation of the model's guidelines and benchmarks, resulting in results that did not accurately reflect an organisation's actual cybersecurity maturity level (Liyanage et al., 2024). Additionally, Dube et al. (2020) and Ghaffari & Arabsorkhi (2018) point out that the broad and all-encompassing nature of Generic CCMMs failed to account for the unique cybersecurity needs of different organisations. This lack of specificity could lead organisations to focus on areas of lesser relevance to their operational context, diverting resources away from critical vulnerabilities unique to their organisation.

In response to these challenges, researchers have explored Specific CCMMs (Akinsanya et al., 2019; Alayo et al., 2021), focusing on the capability domains most relevant to particular sectors or areas of cybersecurity, such as Cloud Security (Le and Hoang, 2017) or Software Security (Akinsanya et al., 2019). By narrowing their focus, these models offered insights tailored to the operational realities and specific threats faced by each sector (Rabii et al., 2020). However, these models often lacked comprehensive coverage of all essential capability domains, leaving organisations exposed to risks outside their primary focus (Aliyu, et al., 2020). Therefore, a holistic perspective is needed to ensure a balanced approach that addresses diverse security needs, even within the same sector.

Several studies have highlighted that existing CCMMs often do not align well with the specific contexts in which organisations operate. Rabii et al. (2020) highlight challenges faced by smaller organisations when applying broadly scoped models that demand extensive resources, often beyond what these organisations can reasonably sustain. Conversely, Liyanage et al. (2024) argue that narrowly focused models often lack the flexibility to incorporate other essential domains, such as business continuity and disaster recovery, making them less suitable for organisations with wider security requirements. Therefore, there is a demand for more flexible and customisable models that adapt to changing cyber threat landscapes and diverse organisational requirements.

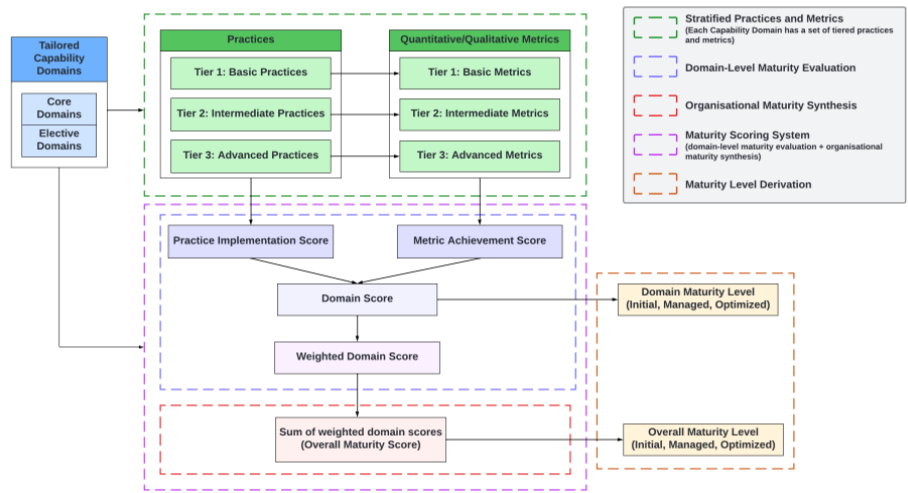

*Figure 1. Overview of the proposed Cybersecurity Capability Maturity Framework.*

Many of the existing CCMMs provide prescriptive guidelines based on regulatory compliance requirements and best practices within the cybersecurity domain (Rea-Guaman et al., 2017). However, they often lack practical tools with automated assessment mechanisms, tailored implementation roadmaps, or integration guidance that make these frameworks easy to apply and embed within diverse organisational workflows (Liyanage et al., 2024). This limits their usability and adoption, especially in resource-constrained environments where organisations may struggle to translate maturity criteria into actionable improvements (Rabii et al., 2020).

Therefore, two key research gaps emerge from the existing literature. First, current CCMMs lack a holistic approach that integrates cybersecurity capabilities across technological, organisational, operational, and human dimensions while offering the flexibility to accommodate diverse organisational contexts. Second, there is an absence of practical tools that make these frameworks easy to implement for organisations. To address these gaps, we propose a novel CCMF that is holistic, flexible, and measurable by integrating comprehensive cybersecurity capabilities across the aforementioned dimensions, enabling customisation for different organisational contexts, and incorporating quantitative metrics to assess and track cybersecurity maturity. We are also developing an interactive, user-centred web-based tool based on the framework, turning its components into practical features that let organisations assess their maturity, customise the assessment, and receive useful recommendations.

## 3 Framework Overview

A CCMM typically consists of three main components: capability domains, practices, and maturity levels. Our framework builds up on these components to provide a more holistic and flexible approach, as shown in Figure 1. Developed through a DSR approach (Peffers et al., 2007), the CCMF integrates insights from existing CCMMs, cybersecurity frameworks, and maturity assessment models. First, **Tailored Capability Domains,** comprising core and elective cybersecurity areas, are derived from our previous work (Liyanage et al., 2024), which analysed existing CCMMs, examining their core functions and specialised focuses. This ensures coverage of key cybersecurity capability domains while allowing organisations to select the domains most relevant to their operations, providing flexibility and alignment with their needs. Second, **Stratified Practices and Metrics** are introduced to distinguish between basic, intermediate, and advanced practices, helping organisations prioritise and progress in a structured manner. This tiered approach is inspired by the CIS Controls (CIS, 2024), which guide security implementation priorities, and the CSIRT Maturity Framework (ENISA, 2022), which emphasises progressive capability development. The integrated qualitative and quantitative metrics facilitate performance measurement and improvement tracking. Lastly, our **Maturity Scoring System** enhances traditional maturity assessments that focus only on whether cybersecurity practices are implemented (Ghaffari & Arabsorkhi, 2018; Karabacak et al., 2016) by also assessing the effectiveness of those practices (metrics). It calculates both domain-specific and overall maturity levels of the organisation to provide clear, actionable insights into their cybersecurity posture, guiding targeted improvements and strategic planning. The web tool's interfaces are designed to incorporate these components in an intuitive manner, allowing organisations to interactively select maturity tiers, evaluate practices and metrics, and view calculated maturity levels, as shown in Figure 2. Grounded in DSR, the framework's iterative design and validation, based on expert input and empirical assessment, are discussed in Section 4.

### 3.1 Tailored Capability Domains

The framework establishes a two-tiered structure addressing technological, organisational, operational, and human aspects (i.e. holistic approach) of developing cybersecurity capabilities. Technological aspects focus on securing infrastructure, systems, and applications through domains such as *Network Security*, *Application Security*, and *Cloud Security*, which mitigate risks arising from software vulnerabilities, misconfigurations, and emerging cyber threats. Organisational and operational aspects are addressed through domains such as *Cybersecurity Governance*, *Compliance & Legal*, *Incident Response*, and *Business Continuity & Disaster Recovery*, ensuring implementation of structured policies, regulatory adherence, and providing organisations with mechanisms to proactively identify risks and respond to incidents. Human aspects, which are often a critical factor in security breaches, are incorporated through *Cybersecurity Culture, Awareness & Training*, *Communication, Collaboration, & Information Sharing*, fostering an informed workforce capable of recognising and mitigating threats.

To provide a structured and adaptable approach to developing capabilities, the framework distinguishes between core and elective domains. Core domains are derived from common functions emphasised consistently across existing CCMMs and represent foundational security practices essential for any organisation's cybersecurity posture. These domains include *Risk Management, Asset & Configuration Management, Identity & Access Management, Data Security, Incident Response, Cybersecurity Culture, Awareness & Training, and Cybersecurity Governance*. The idea behind these core domains is that some security functions are important for all organisations, irrespective of an organisation's industry or size (Liyanage et al., 2024). For instance, Risk Management is essential for any organisation - a hospital assessing threats to patient data or a university managing access to research systems - as it helps identify vulnerabilities and implement controls to minimise potential security threats. Core domains ensure that basic security measures are always taken care of, creating a solid foundation on which other, more specific measures can be added through elective domains.

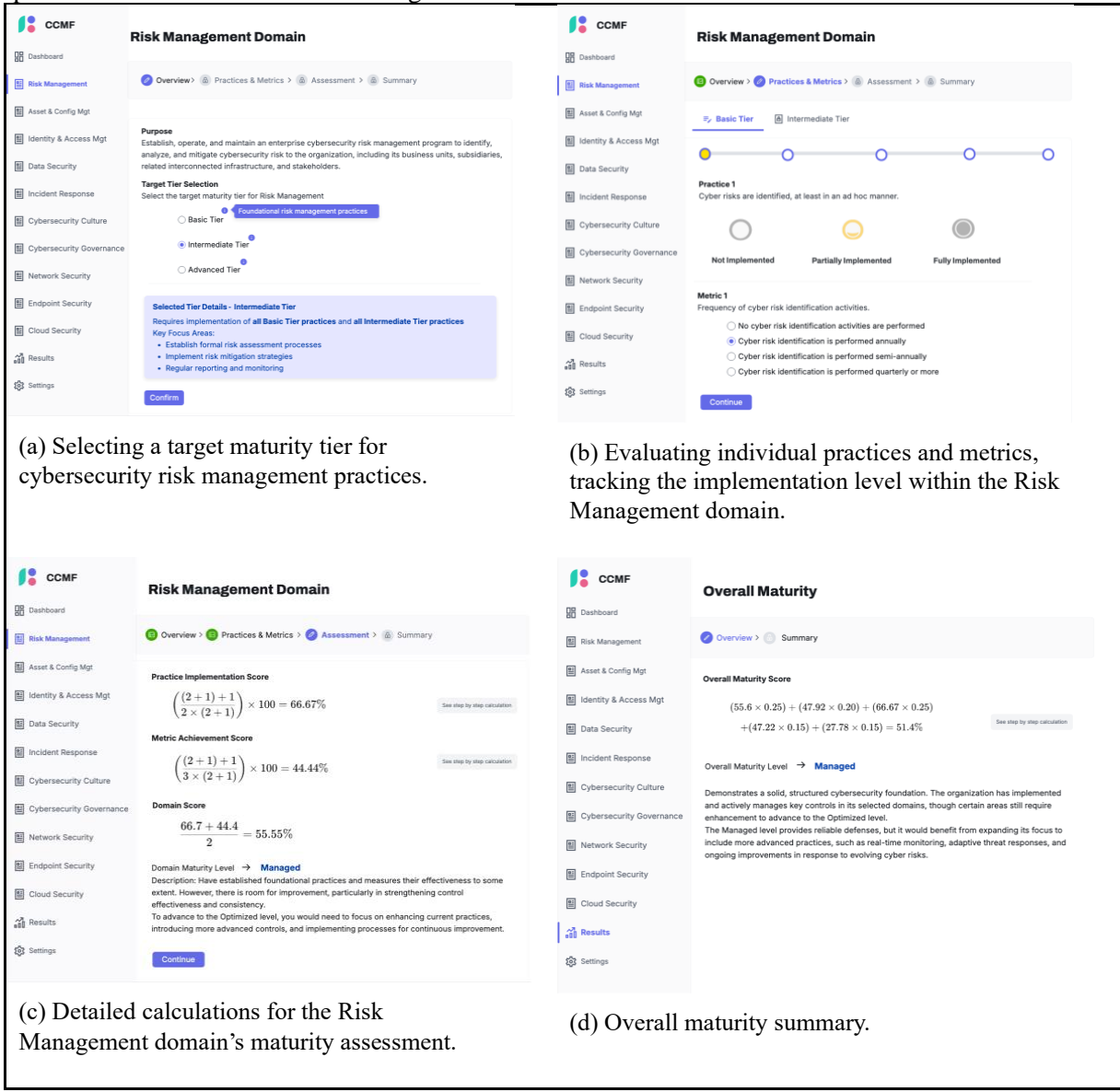

(a) Selecting a target maturity tier for cybersecurity risk management practices.

(b) Evaluating individual practices and metrics, tracking the implementation level within the Risk Management domain.

(c) Detailed calculations for the Risk Management domain's maturity assessment.

(d) Overall maturity summary.

*Figure 2. Sample interfaces of the web-based tool.*

Elective domains are based on specialised security activities identified by analysing the nuanced differences across existing CCMMs. These domains cater to specific organisational needs that may not be universally essential but are critical in certain contexts (Liyanage et al., 2024). Therefore, organisations can choose the elective domains based on their operational context and security priorities. The elective domains include *Network Security, Endpoint Security, Cloud Security, Application Security, Physical Security, Supply Chain & External Dependencies Management, Security Architecture & Design, Situational Awareness, Threat Intelligence & Monitoring, Business Continuity & Disaster Recovery, Workforce Management, Communication, Collaboration, & Information Sharing, Compliance & Legal and Performance Evaluation & Improvement*. For example, technology companies developing cloud-based services may prioritise Cloud Security to ensure the protection of customer data and compliance with industry standards, while manufacturing firms may focus on Supply Chain & External Dependency Management to address cybersecurity risks related to third-party suppliers.

*Novel Cybersecurity Capability Maturity Framework*

## 3.2 Stratified Practices and Metrics

Within each selected domain, our framework assesses maturity through a stratified structure, with each security practice accompanied by tailored metrics to measure progress. This overcomes the issue of misinterpretation and complexity found in previous models that present security practices as a single comprehensive set (Dube and Mohanty, 2020) by breaking down cybersecurity practices into *Basic*, *Intermediate*, and *Advanced* tiers. Each successive tier introduces more complex practices or metrics, building upon the foundational elements established in previous tiers. Organisations need to consider the implementation of practices and achieving metrics from the lower tiers before progressing to higher tiers, ensuring a solid foundation of capabilities. Tiers help organisations prioritise resource allocation, focusing efforts on improving essential practices before tackling more advanced ones. As shown in Figure 2a and Figure 2b, the tool allows users to select their preferred maturity tier for each domain, assess the implementation level of each practice, and track metrics at each tier, facilitating customised assessments that are aligned with specific organisational needs. By incorporating these tiers and measurable criteria, the tool translates principles from capability maturity modelling (Dube and Mohanty, 2020) and dynamic capabilities theory (Naseer et al., 2018) to support structured progression and continuous reassessment as organisational needs, threats, and regulatory environments evolve.

## 3.3 Maturity Scoring System

### 3.3.1 Practice Implementation Score (PIS)

Each practice (*j*) within a domain is evaluated across three levels of implementation, using a predefined Likert scale: "Not Implemented" (assigned 0 points), "Partially Implemented" (assigned 1 point), and "Fully Implemented" (assigned 2 points) (Karabacak et al., 2016), providing a quantitative measure of the extent to which the security practices within a domain have been implemented. For an organisation aiming to achieve a target tier ($t_{target}$), practices from both the target tier and all preceding tiers are considered, ensuring that foundational practices in earlier tiers are implemented before or alongside those in the higher tiers. The Likert scores ($P_{j,t}$) for all practices up to and including the target tier are summed and divided by the product of the cumulative number of required practices ($n_t$) across these tiers and the maximum score (2 points) per practice to get the Practice Implementation Score ($PIS_{t_{target}}$) for the domain.

$$PIS_{t_{target}} = \left( \frac{\sum_{t=1}^{t_{target}} \sum_{j=1}^{n_t} P_{j,t}}{2 \times \sum_{t=1}^{t_{target}} n_t} \right) \times 100 \quad (1) \qquad MIS_{t_{target}} = \left( \frac{\sum_{t=1}^{t_{target}} \sum_{k=1}^{m_t} PE_{k,t}}{3 \times \sum_{t=1}^{t_{target}} m_t} \right) \times 100 \quad (2)$$

### 3.3.2 Metric Achievement Score (MAS)

Metric Achievement Score (MAS) assesses the degree to which the corresponding metrics for those practices have been achieved. To evaluate metric achievement, each metric (*k*) across different tiers (*t*) is assessed based on its performance against pre-defined targets by assigning points ($PE_{k,t}$) on a 0-3 scale. Quantitative metrics use numerical thresholds (e.g. systems with encryption implemented on at least 90% of assets = 3 points, on 70–89% = 2 points), while qualitative metrics rely on rubric-based criteria (e.g. employee understanding of cybersecurity responsibilities = 3 points for clear understanding, 2 points for partial understanding) to ensure consistent evaluation.

MAS (Equation 2) for a domain targeting a specific tier ($t_{target}$) is derived by considering the points earned from all metrics ($m_t$) across the necessary tiers and normalising it by the maximum possible points (3 points) within the tiers, ensuring that each metric's contribution to the overall score is proportionate to its potential impact. Evaluating these practices and metrics through the tool provides organisations with immediate feedback on their current implementation and metric achievement levels and step-by-step calculations to clarify how maturity scores are derived, as shown in Figure 2c.

$$DS_i = \frac{PIS_{t_{target}} + MIS_{t_{target}}}{2} \quad (3) \qquad Maturity\ Level = \begin{cases} Initial, if\ 0 \leq Score \leq 33 \\ Managed, if\ 33 < Score \leq 66 \\ Optimized, if\ 66 < Score \leq 100 \end{cases} \quad (4)$$

*Thirty-Third European Conference on Information Systems (ECIS 2025), Amman, Jordan*      6

### 3.3.3 Domain Score (DS) and Overall Maturity Score (OMS)

By combining the PIS and MAS, the Domain Score ($DS_i$) is calculated to provide a quantitative measure of the capability maturity of the domain (*i*) (Equation 3). A weighted maturity score is calculated for each domain to determine the overall organisational maturity level (Organisational Maturity Synthesis). Weights for each domain ($w_i$) are defined based on their relative importance to the organisation's overall security posture using the Weighted Sum Model (Mohamed et al., 2024). The tool provides an interface where cybersecurity practitioners can assign scores (1–3) to each domain based on four factors: Risk Impact, Compliance Requirement, Business Impact, and Interdependency, aligning with NIST CSF's focus on risk assessment, regulatory compliance, business resilience, and security interdependencies (Almuhammadi & Alsaleh, 2017). It then calculates total scores and normalises them against the sum of all domain scores to generate the final weights. The weighted domain score is calculated by multiplying the $DS_i$ for each domain by its corresponding weight. To get the Overall Maturity Score (Equation 5), the summation of the weighted domain scores of all the selected domains is considered.

$$OMS = \sum_{i=1}^{n}(w_i \times DS_i) \qquad (5)$$

A detailed illustration of how the maturity scoring system is applied in practice, including step-by-step calculations, can be found in supplementary material at https://cutt.ly/craxkJ88.

### 3.3.4 Assigning Maturity Level

The framework defines three levels of maturity: *Initial*, *Managed*, and *Optimized* (Le and Hoang, 2017). At the *Initial* level, basic cybersecurity practices are present, but they lack coordination and are not consistently managed. Organisations at this level may have ad-hoc or fragmented security processes. At the *Managed* level, the cybersecurity program is documented and consistently implemented, with essential controls in place and measured for effectiveness. Finally, at the *Optimized* level, the organisation's cybersecurity posture is continuously improved based on risk assessments and performance metrics, with advanced controls integrated into the overall security strategy.

To assign domain maturity level, $DS_i$ is compared against predefined thresholds given in Equation 4 (Le and Hoang, 2017). Similarly, OMS is compared against the pre-defined ranges to determine the overall cybersecurity maturity level. This completes the Maturity Level Derivation process (Figure 2d). Additionally, the tool includes visual representations, such as graphs and charts, illustrating maturity levels across various domains, allowing organisations to compare performance in different domains easily. These visual aids enable users to identify strengths and weaknesses in specific domains, supporting targeted improvements and strategic planning to enhance cybersecurity maturity.

## 4 Conclusion and Future Work

In this paper, we propose a novel Cybersecurity Capability Maturity Framework designed using DSR to address the limitations of existing models. It provides a tailored and holistic approach for organisations to assess their cybersecurity posture more effectively and align improvement strategies with their unique operational contexts. The introduction of a quantitative maturity scoring system allows for clear, measurable insights into the current maturity level, offering a reliable basis for tracking progress and prioritising cybersecurity enhancements. This framework fills a gap in the current landscape of CCMMs, offering a more flexible solution for organisations seeking to enhance their cybersecurity maturity.

Future work will focus on validating and refining this framework through three main steps. First, an expert review using the Delphi Technique (Karabacak et al., 2016) will empirically investigate the framework, gathering multiple rounds of feedback from security professionals affiliated with organisations such as NCSC and CISA on its clarity and effectiveness in assessing organisational cybersecurity capability maturity. The feedback obtained through this iterative process will be used to refine and improve the framework. Second, the free, web-based tool developed based on the framework will be validated through user testing with participants across diverse organisational sizes and industries (e.g., SMEs, large enterprises, healthcare, finance, critical infrastructure) to assess its usability and effectiveness (Reynolds et al., 2021). Based on the findings from user testing, the tool will be further refined to address identified issues. Finally, a field study (Dube and Mohanty, 2020) will be conducted within real-world organisational settings to assess the tool's effectiveness in accurately evaluating cybersecurity capabilities, with feedback gathered on its usability and value for user adoption.